\documentclass[%
 reprint,
 superscriptaddress,
nofootinbib,
 amsmath,amssymb,
 aps,
 prd,
twocolumn,
notitlepage
]{revtex4-1}
\usepackage{graphicx}
\usepackage{dcolumn}
\usepackage{subfigure}
\usepackage{bm}
\usepackage{breakurl}
\usepackage{enumitem}
\usepackage{xcolor}

\newcommand{\llambda}{\tilde{\lambda}}

\begin{document}


\title{The fate of $O(N)$ multi-critical universal behaviour}

\author{Nicol\`o Defenu}
\email{ndefenu@phys.ethz.ch}
\affiliation{Institut f\"ur Theoretische Physik, Universit\"at 
Heidelberg, D-69120 Heidelberg, Germany}
\affiliation{Institute for Theoretical Physics, ETH Z\"urich
Wolfgang-Pauli-Str. 27, 8093 Zurich, Switzerland}

\author{Alessandro Codello}
\affiliation{Instituto de F\'isica, Faculdad de Ingenier\'ia, Universidad de la Rep\'ublica, 11000 Montevideo, Uruguay}

\begin{abstract}
The multi-critical fixed points of $O(N)$ symmetric models cease to exist in the $N\to\infty$ limit, but the mechanism regulating their annihilation still presents several enigmatic aspects. Here, we explore the evolution of high-order multi-critical points in the $(d,N)$ plane and uncover a complex mosaics for their asymptotic behaviour at large $N$. This picture is confirmed by various RG approaches and constitutes a fundamental step towards the full comprehension of critical behaviour in $O(N)$ field theories. \end{abstract}

\maketitle

 \emph{Introduction:} The study of critical phenomena constitutes one of the major challenges of modern physics and is relevant to a wide range of physical applications, including cell membranes\,\cite{Machta2012}, turbulence\,\cite{Canet2016}, fracture and plasticity\,\cite{Kardar1998, Shekhawat2013}, epidemics\,\cite{Cardy1985}, as well as the celebrated thermodynamic and quantum phase transitions\,\cite{Stanley1987,Sachdev:404196}. The characteristic property of critical systems is scale invariance, which leads to the appearance of power law behaviours for the thermodynamic quantities close to the transition point.

 The privileged tool for understanding emergent scale invariance is renormalization group (RG). In its original form the RG coarse grains and rescales the system yielding ordinary differential equations, which describe the evolution of model as a function of the system scale $k$. Therefore, the scale invariant regions of the phase diagram will appear as fixed points of the RG flow and any system close to criticality will flow near to these fixed points\,\cite{Wilson1975}. The possibility to map the RG description of the system on a set of differential equations in the abstract Hamiltonian space, allows to build a unified description of seemingly different systems, using the  know classifications of dynamical systems\,\cite{Wiggins1990}. The correspondence between the thermodynamic/quantum critical points and the fixed points of dynamical systems justifies the observation of the same power law scaling for thermodynamic functions across very different systems, i.e. the celebrated \emph{universality} phenomenon\,\cite{cardy1996}. 
 
 As RG imposed as the ideal tool to comprehend complex systems, novel paradigms emerged for its implementation, where the RG equations describe the flow of entire thermodynamics functions instead of simple parameters\,\cite{Polchinski1984,Wegner1973,Wetterich1993}.
The current understanding of critical phenomena is rooted on the study of prototypical models for spontaneous symmetry breaking, where the universal behaviour can be explored both as a function of the dimension $d$ and of a, possibly continuous, symmetry index $N$. The $O(N)$ field theories describe a vector order parameter with $N$ components $\boldsymbol{\varphi}$, whose ground state value may be either $O(N)$ symmetric $|\boldsymbol{\varphi}_{0}|=0$ or spontaneously broken $|\boldsymbol{\varphi}_{0}|\neq0$. 

The study of $O(N)$ symmetric models has a long history within RG, which starts with traditional perturbative approaches\,\cite{Wilson1974, Brezin1976, Brezin1993, Moshe2003}, goes through real space and variational results\,\cite{Efrati2014,Kleinert2001, Zinn-Justin1996} and ends up with the comprehensive picture provided by the functional RG approach\,\cite{Codello2013, Codello2015}. In integer dimensions $d\in\mathbb{N}$, the RG picture can be verified with numerically exact results for the universal critical exponents, obtained by MC simulations\,\cite{Pellissetto2002} or conformal bootstrap\,\cite{El-Showk2012}. Interestingly enough most of the RG predictions\,\cite{Zinn-Justin1996,Codello2013, Codello2015} can be straightforwardly extended to real dimension and symmetry index $(d,N)\in\mathbb{R}^{2}$. Far from being simply mathematical speculations, $O(N)$ field theories in fractional dimension have been related to the universal behaviour of lattice models on inhomogeneous structures\,\cite{Burioni1999, Burioni1996, Millan2018, Codello2015} and of power-law decaying interactions\,\cite{Joyce1966, Leuzzi2013, Angelini2014, Defenu2015, Defenu:2017dc, Gori2017}. The analytic continuation of the RG predictions for the critical exponents to the real $(d, N)$ plane has been a fundamental ingredient to the understanding of universality\,\cite{Cardy1980, Peled2017}, especially due to the various mappings between exotic $N$ values ($N=\infty,0,-2,\cdots$) and celebrated statistical mechanics models\,\cite{Stanley1968, deGennes1972, Stanley1968, Balian1973, Fisher1973}. 

Yet, our understanding of the $O(N)$ models phase diagram is far from being complete. Most of our knowledge has been obtained expanding the universal quantities in the $(d,N)$ plane around exactly solvable points and the construction of a general picture needs to integrate these diverse predictions\,\cite{Zinn-Justin1996}. In particular, a major inconsistency has been recently noticed in the behaviour of multi-critical fixed points in the $N\to\infty$ limit, leading to the discovery, in dimensions $2<d<4$, of several non-perturbative fixed points, whose existence was not previously known\,\cite{Yabunaka2017,Yabunaka2018}. In the present letter, we are going to argue that the actual fixed point structure of $O(N)$ field theories is far more complex than the existing one and that the mechanism odescribed in Refs.\,\cite{Yabunaka2017,Yabunaka2018} does not generally applies to all multi-critical fixed points.

\emph{O(N) models:} The Hamiltonian of the $O(N)$ symmetric model in the statistical field theory formalism reads
\begin{align}
\label{h_on}
H=\int d^{d}x\left\{\frac{1}{2}(\nabla\boldsymbol{\varphi})^{2}+V(\rho)\right\}
\end{align}
where $d$ is the spatial dimension, $\boldsymbol{\varphi}(x)$ is a $N$ component vector field and $\rho=\boldsymbol{\varphi}\cdot\boldsymbol{\varphi}/2$ is the scalar order parameter. The traditional perturbative picture is obtained by a Taylor expansion of the local potential $V(\rho)=\mu\rho+\sum_{i=2}^{n}g_{i}\rho^{i}/i!$, where $\mu$ is referred to as the mass and $g_{i}$ are the couplings. The traditional approach to study latter models is to write renormalization group (RG) equations for the couplings $g_{i}$\,\cite{Wilson1975}. Within this approach the conventional symmetry breaking transition which occurs in ferromagnets, superfluids and superconductors in $d=3$ is described by the so called Wilson-Fisher (WF) fixed point. This fixed point is attractive in all directions in the couplings phase space but one, the mass $\mu$, which makes the system off-critical and describes thermal perturbations. The universal properties close to the WF fixed point have been addressed with several different techniques and are known to an exceeding numerical accuracy\,\cite{pelissetto1998,Kleinert2001,ElShowk2012}

As $d$ is lowered, novel \emph{multi-critical} universality classes appear in the perturbative treatment, due to the couplings $g_{i}$ becoming relevant. The series of upper critical dimensions below which the multi-critical fixed points branch out the Gaussian one is $d_{c,n}=2+2/n$, where the index $n=i-1$ labels the universality classes based on the number of infra-red relevant directions at the corresponding fixed point. Then, the $n$th universality class presents $n+1$ universal critical exponent. The numerical computation of the critical exponents for multi-critical universality classes has proved to be challenging both for numerical and perturbative techniques\,\cite{Domb1984}. On the other hand, accurate numerical estimation of these quantities at least for the lowest multi-critical universalities have been possible thanks to the functional RG formalism\,\cite{Codello2013,Codello2015}.

As firstly noticed in Refs.\,\cite{Yabunaka2017,Yabunaka2018}, one major open question regards the fate of the multi-critical universalities in the $N\to\infty$ limit. Specifically, while the $n=1$ fixed point can be analytically continued up to $N=\infty$, where it describes the universal behaviour of the celebrated spherical model\,\cite{Berlin1952, Stanley1968}, no isolated multi-critical fixed point exists at $N\to\infty$ in $d<3$, neither in the field theoretical formalism nor in the lattice description\,\cite{Katsis2018}. Using non-perturbative expansions for the effective action of $O(N)$ models it has been possible to follow the evolution of the first multi-critical fixed point ($n=2$) in $d<3$, showing the existence of an upper critical threshold $N^{2}_{c}(d)$, where the $n=2$ fixed point merges with a non-perturbative partner and disappears from the phase diagram. 

A similar mechanism, where a non trivial solution of the RG equations developed a finite imaginary part a disappeared from the phase diagram at a finite $N$ value, was already noticed in the perturbative treatment of the $n=2$ point\,\cite{Domb1984}, but it was at first associated to an artefact of perturbation theory. On the contrary, the perturbative  coalescence point has been shown to coincide with the non-perturbative threshold for $N^{2}_{c}(d)$\,\cite{Osborn2018}. The picture delineated by previous studies is, therefore, rather straight, as it implies the same mechanism of coalescence for all multi-critical fixed points at large $N$ and it may, at first sight, appear to be complete.

 Actually, this is not the case and the complete scenario describing the \emph{fate of multi-critical universalities at large $N$} is far more complex and it distinguishes between points with an odd or even number of relevant directions.

\emph{Perturbative analysis:} the proof of the statement above can be readily obtained by the perturbative study of higher order multi-critical universalities. It is convenient to briefly recapitulate the result of the $i=3$ case pursued in Ref.\,\cite{Osborn2018}, as the perturbative expressions in this case are relatively simple and the analysis will be more clear. The study of the $i=4,5$ case, necessary to complete the picture, will be outlined in the following and the details reported in the Appendix. In order to investigate the annihilation mechanism of multi-critical universalities it is necessary to consider the leading order (LO) and next to leading order contributions (NLO) to the flow equations of the effective potential and the wave-function renormalization of $O(N)$ field theories in the $\varepsilon=d_{c,n}-d$ expansion formalism. For $n=2$ one has $d_{c,2}=3$ and the flow equations read
\begin{align}
\label{v_flow_lo}\beta V_{\mathrm{LO}}&=\frac{1}{3}V_{ijk}V_{ijk}\\
\beta V_{\mathrm{NLO}}&=\frac{1}{6}V_{ij}V_{iklm}V_{jklm}-\frac{4}{3}V_{ijk}V_{klmn}V_{ijlmn}\nonumber\\
&-\frac{\pi^{2}}{12}V_{ijkl}V_{klmn}V_{ijmn}-\frac{1}{45}V_{ijklmn}V_{ijklmn}\label{v_flow_nlo}\\
\label{z_flow}\beta Z_{\mathrm{NLO}}&=-\frac{1}{45}V_{ijklmn}V_{ijklmn}
\end{align}
where the subscript indicates derivatives with respect to the components of the field and the sum over repeated indexes is intended. It is worth noting that the flow of the wave-function renormalization only depends on the effective potential derivatives at this order, this important property holds for any index $n$. The $n=2$ fixed point can be characterised in terms of a single coupling $\lambda_{3}$, defined by 
\begin{align}
\label{ansatz3}
V(\phi)=\frac{\lambda_{3}}{48}\,\left(\varphi_{i}\varphi_{i}\right)^{3}.
\end{align}
By substituting the ansatz in Eq.\,\eqref{ansatz3} into the $\beta$-functions in Eqs.\,\eqref{v_flow_lo}, \eqref{v_flow_nlo} and \eqref{z_flow} and introducing scaled variables $\lambda_{i}=k^{D_{\lambda}}\tilde{\lambda}_{i}$, with $D_{\lambda}=(d-2+\eta)i-d\,i$, one obtains
\begin{align}
\label{beta_tric}
\beta \llambda_{\mathrm{LO}}&=-2\varepsilon \llambda_{3}+ (88+12N)\llambda_{3}^{2}\\
\beta \llambda_{\mathrm{NLO}}&=-\Biggl(13216+3432N+12N^{2}+\nonumber\\ &\pi^{2}\left(1360+310N+17N^{2}+\frac{N^{3}}{2}\right)\Biggr)\llambda_{3}^{3},
\end{align}
where the anomalous dimension $\eta$ appearing in the definition of scaled variables has been computed using the flow of the wave-function renormalization $Z$ and already included in the NLO term. At LO the $n=2$ fixed point value of the coupling reads
\begin{align}
\label{LOtrisol}
\llambda_{3}=\frac{\varepsilon}{44+6N}
\end{align}
which, as expected, vanishes in $N\to\infty$ limit. Including next to leading order terms one gets three solutions, an infrared stable one $\lambda^{*}_{\mathrm{IR}
}$, whose leading order is given by Eq.\,\eqref{LOtrisol} and two ultraviolet stable fixed points, the Gaussian fixed point at $\lambda=0$ and a novel fixed point $\lambda^{*}_{\mathrm{UV}}$. As $\varepsilon>0$ grows at fixed $N$ the two non-trivial fixed points $\lambda^{*}_{\mathrm{IR}}$ and  $\lambda^{*}_{\mathrm{UV}}$  approach each other and finally collide at the threshold value
\begin{align}
\label{conv_rad}
\varepsilon_{c}=\frac{36}{\pi^{2}}\frac{1}{N}+O\left(\frac{1}{N^{2}}\right)
\end{align}
which was firstly identified as the convergence radius of $\varepsilon$-expansion\,\cite{Pisarski1982}. The expression in Eq.\,\eqref{conv_rad}  suggests the possibility to recast the large $N$ perturbative $\beta$-functions in terms of the composite variable $N\varepsilon$ with constant convergence radius. This task is readily accomplished by the introduction of a scaled variable $\llambda_{3}\to\Lambda_{3}/N^{2}$, obeying the $\beta$-function
\begin{align}
\label{tric_beta_ln}
\beta\Lambda_{3}=-2\varepsilon\Lambda_{3}+\left(12-\frac{\pi^{2}}{2}\Lambda_{3}\right)\frac{\Lambda_{3}^{2}}{N}+O(N^{-2})
\end{align} 
with three solutions at leading order in $N\varepsilon$ 
\begin{align}
\Lambda_{3}&=0\\
\Lambda_{3}&=\frac{2}{\pi^{2}}\left(6\pm\sqrt{36-N\varepsilon\pi}\right).
\end{align}
Therefore,  Eq.\,\eqref{tric_beta_ln} encodes all the necessary information to derive the collapse of the $n=2$ fixed point already at leading order in $N\varepsilon$. The convergence radius for the composite parameter is $N\varepsilon=(\frac{6}{\pi})^{2}$, as in Eq.\,\eqref{conv_rad}, and it yields the threshold curve for the existence of the $n=2$ fixed point at large $N$ 
\begin{align}
N_{c}^{3}(d)=\left(\frac{6}{\pi}\right)^{2}\frac{1}{3-d}.
\end{align}
Non-perturbative analysis of the evolution of the $i=3$ fixed point at large $N$ produces an analogous picture with the collision between the $i=3$ fixed point and a UV stable non-perturbative partner on a threshold line $N_{c}^{3}(d)\approx 3.6/(3-d)$, demonstrating the reliability of the $\varepsilon$-expansion approach\,\cite{Yabunaka2017, Pisarski1982}. 

\emph{Tetra-critical universality:} the perturbative analysis in the $n=3$ case follows the same lines, but it yields a completely different picture. The expression for the flow of the potential at next to leading order in $\varepsilon=d_{\mathrm{c},3}-d=8/3-d$ is rather cumbersome. Then we will report only the LO here, while the NLO can be found in App.\,A, see also Ref.\,\cite{Zinati2019}
\begin{align}
\label{qd_v_flow_lo}\beta V_{\mathrm{LO}}&=\frac{1}{8}V_{ijkl}V_{ijkl}
\end{align}
where the subscript indicates once again derivatives with respect to the components of the field and the sum over repeated indexes is intended. The flow of the marginal coupling $\lambda_{4}$ for the tetra-critical model is promptly obtained by substituting the ansatz 
\begin{align}
\label{quad_ansatz}
V(\phi)=\frac{\lambda_{4}}{384}\,\left(\varphi_{i}\varphi_{i}\right)^{4}
\end{align}
into the potential flow equations, see Eq.\,\eqref{qd_v_flow_lo} and refer to App.\,A for the NLO. Introducing scaled variables, indicated by the $\sim$ superscript, one obtains the LO $\beta$-function
\begin{align}
\label{beta_qd}
\beta \llambda_{\mathrm{LO}}&=-3\varepsilon \llambda_{4}+ \left(\frac{9}{4}N^{2}+\frac{225}{2}N+804\right)\llambda_{4}^{2}
\end{align}
The $n=3$ fixed point solution derived from Eq.\,\eqref{beta_qd} including the NLO term found in App.\,A vanishes as $N^{-2}$ in the large $N$ limit and indicates the rescaling $\llambda_{4}\to\Lambda_{4}/N^{3}$ as the appropriate one in order to recast the flow equations in terms of the combined parameter $N\varepsilon$, yielding
\begin{align}
\beta\Lambda_{4}=-3\varepsilon\Lambda_{4}+\frac{9\Lambda_{4}^{2}}{4N}+O(N^{-2})
\end{align} 
leading to the two tetra-critical fixed point solutions
\begin{align}
\Lambda_{4}&=0\\
\Lambda_{4}&=\frac{4N\varepsilon}{3}.
\end{align}
Surprisingly these two solutions always exist irrespectively from the $N\varepsilon$ value. At $\varepsilon=0$ both the solutions coincide with the Gaussian fixed point, while at larger $\varepsilon$ the correlated solution branches out from the quadratic one and continues to exist for any finite $N\varepsilon$.  As a consequence, no threshold for the existence of the tetra-critical universality class at large $N$ is found ($N^{3}_{c}(d)=\infty$).

The perturbative analysis evidences a possibly different fate for multi-critical universality in the large $N$ limit. In order to confirm this picture, we have further investigated the evolution of the $n=3$  fixed point using the functional RG approach. Projecting the exact RG equation\,\cite{Polchinski1984,Wegner1973,Wetterich1993} for the effective action on a functional ansatz with the same form of Eq.\,\eqref{h_on}, but with a scale dependent effective potential $V_{k}(\rho)$, one finds the non-perturbative $\beta$-function of the effective potential in the so called local potential approximation (LPA)\,\cite{Berges2002, Delamotte:2011jk}.
\begin{figure*}[ht!]
\centering
\subfigure[\quad $d=2.6$]{\includegraphics[scale=0.3]{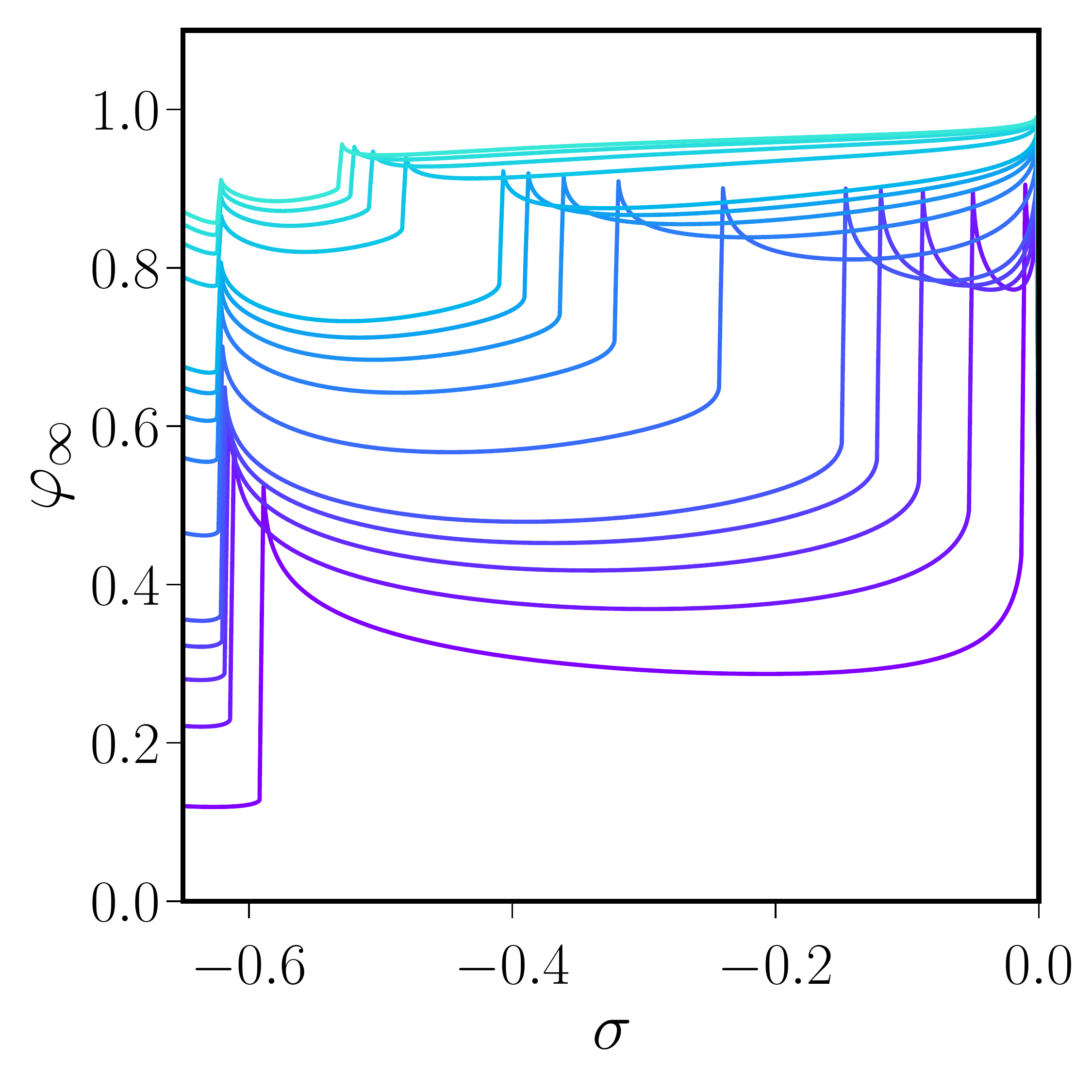}}
\hspace{5mm}
\subfigure[\quad $d=2.45$]{\includegraphics[scale=0.3]{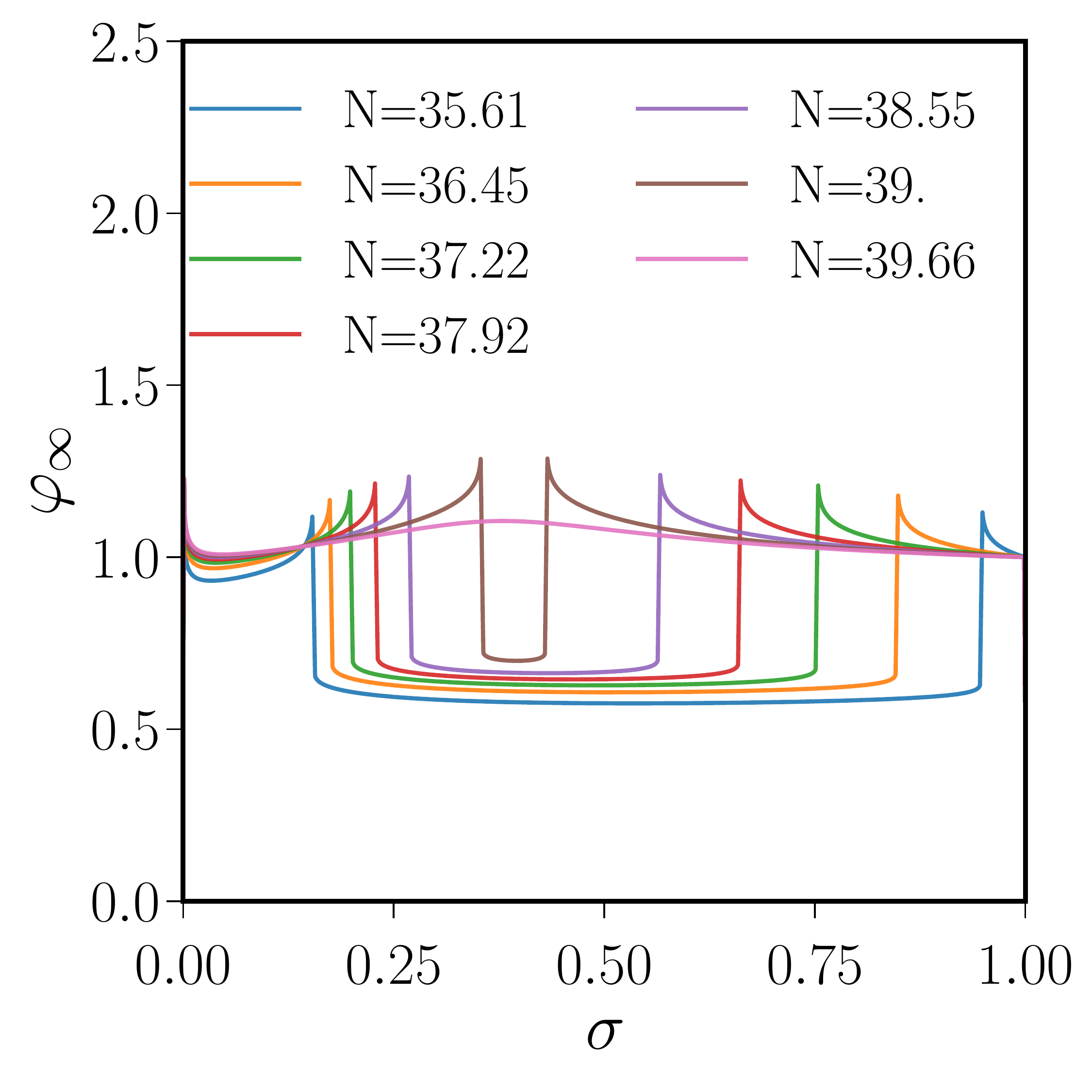}}
\caption{The function $\varphi_{\infty}(\sigma)$ in $d=2.6$ starting at small $N$ (panel a) and at very large $N$ (panel b) for negative and positive $\sigma$ respectively. In panel (a) the leftmost peak is the Wilson-Fisher universality which remains fixed for growing $N$, on the contrary the right peak represent the tetra-critical universality which slowly approaches the WF. This is confirmed by the analysis pursued at very large values of $N$ where the tetra-critical comes closer to the WF universality. \label{Fig1}}
\end{figure*}

The analysis of the LPA flow equation has been pursued numerically using the shooting technique already employed to draw the comprehensive picture of the universal properties for the $n=1$ fixed point in $O(N)$ field theories both with short-range\,\cite{Codello2013, Codello2015,Defenu:2017el, DefenuJHEP} and long-range interactions\,\cite{Defenu2015, Defenu2016, Defenu:2017dc,Defenu2017a}. The shooting method consists in solving the effective potential flow equation as a function of the initial condition $V_{k}''(\rho)=\sigma$, obtaining a numerical estimation of the function $\varphi_{\infty}(\sigma)$, which represents the value of the field at which the effective potential develops a singularity. Then, the physical fixed point potentials will be signalled by a divergence or a singularity of the $\varphi_{\infty}(\sigma)$ function\,\cite{Morris1994, Codello2013, Hellwig2015, Codello2015}. The function $\varphi_{\infty}(\sigma)$ for $d=2.6$  for different values of $N$ is shown in Fig.\,\ref{Fig1} panel (a), surprisingly the annihilation phenomenon previously described for the $n=2$ universality does not apply here and the tetra-critical fixed point simply shifts to smaller $\sigma<0$ values for growing $N$. 

The justification of the discrepancy between the fate of the $n=2$ fixed point and the $n=3$ one cannot be found by any of the arguments presented in Ref.\,\cite{Yabunaka2017, Yabunaka2018} as they generically apply to all multi-critical universality classes on equal footing and, implicitly, suggest a common annihilation mechanism. Yet,  this is not the case and the large $N$ behaviour of the tetra-critical universality presents no analogy with the $n=2$ case. The present analysis points at a more complex scenario for multi-critical universality classes in the large $N$ limit with odd order ($n\in2\mathbb{Z}+1$) fixed points not showing any peculiar annihilation mechanism at large $N$. 

\emph{Penta-critical universality:}
In order to support latter conjecture, we perform the perturbative analysis of the $n=4$ fixed point and show that its behaviour corresponds to the one found in the $n=2$ case. The full analysis of the pentacritical case can be found in App.\,B, here we only report the $\beta$-function for the scaled coupling $\Lambda_{5}=N^{3}\tilde{\lambda}_{5}$
\begin{align}
\label{beta_pc_ln}
\beta &\Lambda_{\mathrm{LO}}=-4 \Lambda_{5}  \epsilon\nonumber\\
&+\frac{160 \Lambda_{5}^{2}+\left(\sqrt{2\pi}\frac{80}{3}\Gamma\left(\frac{1}{4}\right)^{2}+\frac{\sqrt{2}2560}{27\sqrt{\pi}}\Gamma\left(\frac{3}{4}\right)^{2}\right)\Lambda_{5}^{3}}{N}.
\end{align}
As in the tri-critical case, see Eq.\,\eqref{tric_beta_ln}, the $\beta$-function in Eq.\,\eqref{beta_pc_ln} allows for two non-trivial solutions which collide at a threshold value $N^{4}_{c}(d)$, leading to the annihilation of the penta-critical fixed point.
Moreover, also in this case the perturbative expression for the $N^{4}_{c}(d)$ value assumes the form 
\begin{align}
\label{ncd5}
N_{c}^{n}(d)=\frac{a_{n}}{d_{\mathrm{c},n}-d}
\end{align}
with the proportionality coefficient $a_{4}=2.035$ in the penta-critical case.

In order to obtain a numerical estimate for the $N^{4}_{c}(d)$ threshold, we pursued the shooting method for the $n=4$ universality in various dimensions $d\in[2.3,2.5]$, where the lower limit has been chosen in order to avoid complications arising from the appearance of additional multi-critical fixed points. The resulting curve is in excellent agreement with the perturbative result in Eq.\,\eqref{ncd5}. Indeed, by fitting the curve obtained via the shooting method we obtain the numerical estimates $a_{4}=2.02(2)$ and $d_{\mathrm{c},4}=2.500(01)$, which perfectly reproduce the perturbative result. The annihilation mechanism for the $n=4$ fixed point at large $N$ at LPA level is reported in Fig.\ref{Fig1} panel (b); its analogy with the $i=3$ case is evident. 

\emph{Fate of multi-critical fixed points at large $N$:} In conclusion, we found extensive evidences that only multi-critical fixed points with an even number of relevant perturbations vanish in the $N\to\infty$ limit by colliding with additional UV stable fixed points. While odd order multi-critical fixed points remain well defined for all finite $N$ values and do not present any non-perturbative partners.

Furthermore, the previous discussion clarified that a proper large $N$ limit has to be taken at fixed $\alpha=N\varepsilon$ and, thus, each fixed point in the $(d,N)$ plane must have its counterpart at $N\to\infty$ at $d=d_{\mathrm{c},n}$. Applying this procedure to the $n=2$  fixed point and its UV stable partner projects each of them on single point in the Bardeen-Moshe-Bander (BMB) line found at $N=\infty$ in $d=3$\,\cite{Bardeen1984}. Moreover, any fixed point involved in the annihilation mechanism of the $n=2$ case appears to have an $N=\infty$ counterpart in $d=3$\,\cite{Fleming2020}. Repeating this analysis for any $n\in 2\mathbb{N}$ shall yield the same result, as infinite generalisations of the BMB phenomenon are expected to appear at the upper critical dimensions of even order multi-critical universality classes\,\cite{Comellas1997}.

However, latter picture cannot apply to $n\in 2\mathbb{N}+1$ fixed points, as no BMB phenomenon exists at the upper critical dimensions $d_{\mathrm{c},n}$ in this case. Rather, an infinite continuous line of fixed points with Gaussian critical exponents, but non-quadratic effective potential exists, with only the isolated $n=1$ fixed point standing aside\,\cite{Comellas1997}. Therefore, the correlated $N\to\infty$ limit can be taken at any fixed $\alpha$ for $n=3,5,7,\cdots$ without encountering any singular point and smoothly mapping any fixed point in the $(d,N)$ plane to $N\to\infty$ and $d=d_{\mathrm{c},n}$.

This work paves the way for a full comprehension of $O(N)$ symmetric models and yields a fundamental piece of information to understand several open issues in connected critical models. It would be particularly interesting to study how the presently described mechanism generalises to the Fermionic and supersymmetric cases\,\cite{Gehring2015, Hellwig2015,Heilmann2012} and what are the implications of this mechanism on current experiments featuring high-order critical points\,\cite{Zwerger2019}.

\emph{Acknowledgements}: The authors are grateful to B. Delamotte and S. Yabunaka for fruitful discussions. This work is supported by the Deutsche Forschungsgemeinschaft (DFG, German Research Foundation) via Collaborative Research Centre “SFB1225” (ISOQUANT) and under Germany’s Excellence Strategy “EXC-2181/1- 390900948” (the Heidelberg STRUCTURES Excellence Cluster).
\onecolumngrid
\appendix
\section{tetracritical flow equations}
The vertex flow equations  at next-to-leading order in the functional $\varepsilon$-expansion around the critical dimension $d_{c}=8/3$ read
%
\begin{align}
\label{qd_v_flow_lo}\beta V_{\mathrm{LO}}&=\frac{1}{8}V_{ijkl}V_{ijkl}\\
\beta V_{\mathrm{NLO}}&=\frac{1}{160}V_{ij}V_{iklmnop}V_{jklmnop}
-\frac{9}{80}V_{ijk}V_{klmnop}V_{ijlmnop}
-\frac{3}{8}V_{ijkl}V_{kmnop}V_{ijlmnop},\nonumber\\
&-\frac{\Gamma\left(\frac{1}{3}\right)^{3}}{24}V_{ijklmn}V_{ijkop}V_{lmnop}
\frac{9}{64}\left(\frac{\sqrt{3}\pi}{3}-2+\log(3)\right)V_{ijkl}V_{ijmnop}V_{klmnop},\label{qd_v_flow_nlo}\\
\label{qd_z_flow}\beta Z_{\mathrm{NLO}}&=-\frac{1}{1120}V_{ijklmnop}V_{ijklmnop}.
\end{align}
%
Introducing the following ansatz for the effective potential at the tetra-critical fixed points
\begin{align}
\label{quad_ansatz}
V(\phi)=\frac{\lambda_{4}}{384}\,\left(\varphi_{i}\varphi_{i}\right)^{4}
\end{align}
into the potential and wave-function flows Eqs.\,\eqref{qd_v_flow_lo}, \eqref{qd_v_flow_nlo} and \eqref{qd_z_flow} one obtains the $\beta$-function for the coupling $\lambda_{4}$
\begin{align}
\label{beta_qd}
\beta \llambda_{\mathrm{LO}}&=-3\varepsilon \llambda_{4}+ \left(\frac{9}{4}N^{2}+\frac{225}{2}N+804\right)\llambda_{4}^{2}\\
\beta \llambda_{\mathrm{NLO}}&=-\Biggl(\frac{81}{8}N^{4}+\frac{9315}{8}N^{3}+\frac{81081}{2}N^{2}+\frac{942435}{2}N
+1734750-\sqrt{3}\pi\Bigl(\frac{27}{16}N^{4}+108N^{3}+\frac{7641}{2}N^{2}\nonumber\\
&+\frac{90909}{2}N+167670\Bigr)-\Gamma\left(\frac{1}{3}\right)^{3}\Bigl(90N^{3}+3816N^{2}
+50160N+203184\Bigr)-\log(3)\Bigl(\frac{81}{16}N^{4}+324N^{3}\nonumber\\
&+\frac{22923}{2}N^{2}+\frac{272727}{2}N+503010\Bigr)\Biggr)\llambda_{4}^{3}.
\end{align}
Apart for the trivial Gaussian solution $\llambda_{4}=0$, the LO $\beta$-function supports the fixed point solution
\begin{align}
\llambda_{4}=\frac{4\varepsilon}{1072+150N+3N^{2}}
\end{align}
in analogy with the tri-critical case this solution vanishes in the large $N$ limit. However, in this case, in order to make the solution finite, one has to introduce the rescaling
\begin{align}
\llambda_{4}=\frac{\Lambda_{4}}{N^{3}}
\end{align}
leading to the large $N$ fixed point solution discussed in the main text, see Eq.\,(17).
\section{Pentacritical flow equations}
The vertex flow equations for the $\varepsilon$-expansion around the upper critical dimension $d_{\mathrm{uc}}=5/2$ read,
\begin{align}
\label{penta_vlo}
\beta V_{\mathrm{LO}}&=\frac{1}{30}V_{ijklm}V_{ijklm},\\
\label{penta_vnlo}
\beta V_{\mathrm{NLO}}&=\frac{1}{7560}V_{ij}V_{iklmnopqr}V_{jklmnopqr}-\sqrt{\frac{\pi}{2}}\frac{\Gamma\left(\frac{1}{4}\right)^{2}}{144}V_{ijklmnop}V_{ijklqr}V_{mnopqr}+\frac{\Gamma\left(\frac{3}{4}\right)^{2}}{\sqrt{2\pi}45}V_{ijkl}V_{ijmnopqr}V_{klmnopqrs}\nonumber\\
&-\frac{\pi\Gamma\left(\frac{1}{4}\right)^{2}}{216\Gamma\left(\frac{3}{2}\right)^{2}}V_{ijklmn}V_{ijkopqr}V_{lmnopqr}+\frac{2}{945}V_{ijk}V_{ilmnopqr}V_{jklmnopqr}+\frac{2}{135}V_{ijkl}V_{imnopqr}V_{jklmnopqr}\nonumber\\
&-\frac{2}{45}V_{ijklm}V_{inopqr}V_{jklmnopqr}
+\frac{\pi-4-\log(4)}{45}V_{ijklm}V_{ijnopqr}V_{klmnopqr},\\
\beta Z_{NLO}&=-\frac{1}{56700}V_{ijklmnopqr}V_{ijklmnopqr}\label{penta_znlo}\end{align}
Once again, we formulate an ansatz for the effective potential at the penta-critical fixed point \begin{align}
\label{p_ansatz}
V(\phi)=\frac{\lambda_{5}}{720}\,\left(\varphi_{i}\varphi_{i}\right)^{5},
\end{align}
and insert it into the potential and wave-function flows, Eqs.\,\eqref{penta_vlo},\,\eqref{penta_vnlo} and\,\eqref{penta_znlo} we obtain the $\beta$-function for the coupling
\begin{align}
\label{beta_qd}
\beta \llambda_{\mathrm{LO}}&=-4 \llambda_{5}  \epsilon+\llambda_{5} ^2 \left(160 N^2+\frac{15680 N}{3}+\frac{110848}{3}\right)\\
\beta \llambda_{\mathrm{NLO}}&=-\llambda_{5} ^3 \Biggl(\frac{7111168 N^4}{81}+\frac{486225920
   N^3}{81}+\frac{11685017600 N^2}{81}+\frac{117273763840 n}{81}+\frac{139819679744}{27}\nonumber\\
  & -\pi  \left(\frac{51200 N^4}{3}+\frac{3645440 N^3}{3}+\frac{268943360 N^2}{9}+303185920
   N+\frac{9793699840}{9}\right)\nonumber\\
   &+\log (4)\left(-\frac{51200 N^4}{3}-\frac{3645440 N^3}{3}-\frac{268943360 N^2}{9}-303185920
   N-\frac{9793699840}{9}\right) \nonumber\\
   &+ \Gamma \left(\frac{1}{4}\right)^2\left(-15360 N^4-\frac{11018240 N^3}{9}-\frac{306872320 N^2}{9}-\frac{3501301760
   N}{9}-\frac{13961789440}{9}\right)\nonumber\\
   &+\sqrt{\frac{2}{\pi }}\Gamma \left(\frac{3}{4}\right)^2 \left(\frac{2560
   N^5}{27}+\frac{81920 N^4}{9}+\frac{13127680 N^3}{27}+\frac{90337280 N^2}{9}+\frac{2318172160
   N}{27}+\frac{2333081600}{9}\right)\nonumber\\
   & +\sqrt{2 \pi }\Gamma
   \left(\frac{1}{4}\right)^2 \left(-\frac{80 N^5}{3}-\frac{9280
   N^4}{3}-\frac{616640 N^3}{3}-\frac{15819520 N^2}{3}-\frac{509219840 N}{9}-\frac{1941913600}{9}\right) \Biggr).
\end{align}
The LO term for the coupling flow, Eq.\,\eqref{beta_qd}, is sufficient to obtain the fixed point solution for the penta-critical fixed point
\begin{align}
\label{penta_fp}
\llambda_{5}=\frac{3\varepsilon}{8(3464+490N+15N^{2}).}
\end{align} 
Analogously with the tetra-critical case one  can introduce the rescaling
\begin{align}
\llambda_{5}=\frac{\Lambda_{5}}{N^{3}},
\end{align}
which produces a finite large $N$ fixed point. However in this case the physical picture obtained differs from the tetra-critical case, while it is qualitatively identical to the tricritical one, see the main text.
\twocolumngrid
\bibliography{./bibliography.bib}
\end{document}